\documentclass[doublecol]{epl2}
\pdfoutput=1
\usepackage{amssymb}
\newcommand{\half}{\frac{1}{2}}
\newcommand{\thrd}{\frac{1}{3}}
\newcommand{\frth}{\frac{1}{4}}
\newcommand{\sxth}{\frac{1}{6}}
\newcommand{\dar}{\partial_r}
\newcommand{\R}{{\bar r}}
\title{Conformal theory of gravitation and cosmology}
\author{R. K. Nesbet}
\institute{
IBM Almaden Research Center,
650 Harry Road,
San Jose, CA 95120-6099, USA}
\date{\today}
\pacs{04.20.Cv}{Fundamental problems and general formalism}
\pacs{98.80.-k}{Cosmology}
\pacs{11.15.-q}{Gauge field theories}
\abstract{
The postulate of universal local Weyl scaling (conformal) symmetry 
modifies both general relativity and the Higgs scalar field model.
Conformal gravity (CG) has recently been fitted to rotation data for 
138 galaxies. The conformal Higgs model (CHM) acquires a gravitational 
effect that fits observed Hubble expansion for redshifts $z\leq 1$ 
(7.33 Gyr) accurately with only one free constant parameter.
Conformal theory explains dark energy and does not require dark matter, 
providing a viable alternative to the $\Lambda$CDM standard paradigm.  
Vanishing of centripetal acceleration outside a galactic halo boundary 
is a unique implication of the theory. 
The fit to observed astrophysical data by conformal models CG and CHM 
is shown here to account for both parameters 
$w^2$ and $\lambda$ of postulated Higgs potential
$V=-(w^2-\lambda\Phi^\dag\Phi)\Phi^\dag\Phi$, responsible for 
symmetry-breaking finite $\Phi^\dag\Phi$ in particle theory. 
Recent criticism of CG is resolved here by showing that CG and CHM 
are interdependent but compatible.  Nonclassical CG radial acceleration 
parameter $\gamma$ is determined by the CHM.  A recently established 
empirical relationship between classical and nonclassical 
galactic radial acceleration requires parameter $\gamma$ 
to be independent of galactic mass.  Conformal theory is shown here 
to be consistent with this and with the $v^4$ baryonic Tully-Fisher 
relation for galactic rotation velocities.}
\begin{document}
\maketitle

\section{Introduction}
\par In the consensus $\Lambda$CDM paradigm for cosmology, gravitational
phenomena that cannot be explained by general relativity as formulated 
by Einstein are attributed to cold dark matter. The search for tangible 
dark matter has continued for many years\cite{SAN10}, without result. 
Dark energy $\Lambda$ remains without an explanation.
\par Consideration of an alternative paradigm is motivated by this
situation.  An alternative postulate is that of universal conformal
symmetry, requiring local Weyl scaling covariance 
\cite{WEY18,MAN06,MAN07} for all massless elementary physical fields, 
without dark matter\cite{NES13}.
Conformal symmetry, valid for fermion and gauge boson fields
\cite{DEW64}, is extended to both the metric tensor field of
general relativity and the Higgs scalar field of elementary-particle
theory \cite{HIG64,CAG98}, without any novel elementary fields. 
This postulate is exemplified by conformal gravity 
CG\cite{MAK89,MAK94,MAN90,MAN91,MAN06,MAN12} and by the 
conformal Higgs model CHM\cite{NESM1,NESM2,NESM3}.
\par Conformal gravity (CG)\cite{MAN06} retains the logical structure
of general relativity, but replaces the Einstein-Hilbert Lagrangian
density by a quadratic contraction of the conformal Weyl
tensor\cite{WEY18}.  Conformal gravity ensures consistency of the
gravitational field equations, while preserving subgalactic
phenomenology\cite{MAK89,MAN06}.  The conformal Higgs model (CHM)
introduces a gravitational term confirmed by observed Hubble 
expansion\cite{NESM1,NESM2}.
\par Substantial empirical support for this proposed break with
convention is provided by recent applications of CG to 
galactic rotation velocities
\cite{MAN97,MAO11,MAO12,OAM12,OAM15} and of the CHM
to Hubble expansion \cite{NESM1,NESM2}, in a consistent model of 
extended dark galactic halos\cite{NESM3}.
As recently reviewed \cite{MAN06,MAN12,NES13},
conformal theory fits observed data for an isolated galaxy 
without invoking dark matter, resolving several longstanding 
paradoxes.  Revision of theory for galaxy formation 
and galactic clusters is implied \cite{NES13,NESM3}.
\par CG and the CHM are not obviously compatible. The CHM supports 
the Higgs mechanism, spontaneous SU(2) symmetry-breaking, which also 
breaks conformal symmetry\cite{NESM2} and invalidates a transformation 
connecting the two distinct metrics invoked for successful fits to 
galactic rotation and Hubble expansion.
The present paper is concerned with reconciling CG and the CHM in a
valid universal conformal theory\cite{NES13}.
\par The argument here examines the interdependence of CG and the CHM
in the context of a depleted dark halo model \cite{NESM3} of an 
isolated galaxy.  A galaxy of mass $M$ is modeled by spherically
averaged mass density $\rho_G/c^2$ within $r_G$, formed by
condensation of primordial uniform, isotropic matter of mass
density $\rho_m/c^2$ from a sphere of large radius $r_H$ \cite{NESM3}.
The dark halo inferred from gravitational lensing and centripetal
acceleration is identified with this depleted sphere \cite{NESM3}.
A model valid for nonclassical gravitation can take advantage of
spherical symmetry at large galactic radii, assuming classical
gravitation within an effective galactic radius $r_G$.  Non-spherical
gravitation is neglected outside $r_G$. Given mean mass density
${\bar\rho}_G/c^2$ within $r_G$, this implies large empty halo radius
$r_H=r_G({\bar\rho}_G/\rho_m)^\thrd$.
\par CG, CHM, and the depleted halo model are shown here to be mutually
consistent and to agree with observed phenomena including excessive 
rotation velocities and accelerating Hubble expansion.  This resolves 
open questions regarding implications of conformal gravity
\cite{FLA06,BAV09,YOO13}, including the possibility that the CHM might
cancel out the nonclassical acceleration of CG \cite{KH016}.

\section{Overview}
\par CG and CHM must be consistent for an isolated galaxy and its
dark halo, observed by gravitational lensing.  CG is valid for anomalous
outer galactic rotation velocities in the static spherical Schwarzschild
metric, solving a differential equation for Schwarzschild gravitational
potential $B(r)$ \cite{MAK89,MAN06}.  The CHM is valid for cosmic
Hubble expansion in the uniform, isotropic FLRW metric, solving a
differential equation for Friedmann scale factor $a(t)$ \cite{NESM1}.
Concurrent validity is achieved here by introducing a common hybrid
metric.  The two resulting gravitational equations are decoupled by
separating mass/energy source density $\rho$ into uniform isotropic
mean density ${\bar\rho}$ and residual ${\hat\rho}=\rho-{\bar\rho}$,
which extends only to galactic radius $r_G$ and integrates to zero
over the defining volume.
\par The unique Lagrangian density ${\cal L}_g$ of conformal gravity
theory, constructed from the conformal Weyl
tensor \cite{WEY18,MAK89,MAN06}, determines source-free
Schwarzschild gravitational potential
\begin{eqnarray} \label{Brfn}
B(r)=-2\beta/r+\alpha+\gamma r-\kappa r^2,
\end{eqnarray}
valid outside a spherically symmetric mass/energy source density
\cite{MAK89,MAN91}. This adds two constants of integration to the 
classical external potential: nonclassical radial acceleration 
$\gamma$ and halo cutoff parameter $\kappa$ \cite{NESM3}. 
\par
A particular solution for $B(r)$ \cite{MAK89}, assumed by subsequent
authors, derives an integral for $\gamma$ that vanishes for residual
source density ${\hat\rho}$.  This is replaced here by an alternative
particular solution for which $\gamma$ is a free parameter.  The two
solutions differ by a boundary condition at the coordinate origin.
The depleted halo model \cite{NESM3} assumes source-free region 
$r_G\leq r\leq r_H$.   The halo model determines centripetal $\gamma$
from the incremental cosmic radial acceleration due to absence
of cosmic background mass-energy $\rho_m$ in the halo \cite{NESM3},
using the proposed alternative solution of the Schwarzschild equation
for $B(r)$.  This implies $\gamma$ independent of galactic mass,
consistent with a recent study of rotation velocities for 153
galaxies with directly measured mass values \cite{MLS16,NESM7}.
\par Higgs $V(\Phi^\dag\Phi)=-(w^2-\lambda\Phi^\dag\Phi)\Phi^\dag\Phi$
introduces two assumed constants $w^2$ and $\lambda$\cite{HIG64,CAG98}.
In the conformal Higgs model, unique Lagrangian density
${\cal L}_\Phi$\cite{MAN06,NESM1,NESM2} of Higgs scalar field
$\Phi$ adds a gravitational term to $-V$, so that
$(\partial_\mu\Phi)^\dag\partial^\mu\Phi$ is augmented by
\begin{eqnarray} \label{DeltaL}
\Delta{\cal L}_\Phi=
 (w^2-\sxth R-\lambda\Phi^\dag\Phi)\Phi^\dag\Phi.
\end{eqnarray}
This differs from Ref\cite{MAN06}, which identifies Higgs parameter 
$w^2$ with $-\sxth R$. 
Scalar $R=g_{\mu\nu}R^{\mu\nu}$ is the trace of the Ricci tensor.
Assumed nonzero $w^2$ and $\lambda$ are not determined 
by standard theory \cite{HIG64,CAG98}. Concurrent solution of conformal 
scalar and metric tensor field equations determines nonzero
$\Phi^\dag\Phi=\phi^2$ (the Higgs mechanism) and replaces Einstein
tensor $G^{\mu\nu}$ in the gravitational equation by traceless
conformal tensor $R^{\mu\nu}-\frth g^{\mu\nu}R$.
In contrast to \cite{MAN06}, CHM is developed here as a theory formally
independent of CG.  Observed data of Hubble expansion, described by
CHM, determine values of parameters defined by CG. 
The two theories are shown to be compatible.
\par The Higgs mechanism implies a dynamical origin of $w^2$, dark energy 
in the conformal Friedmann cosmic evolution equation \cite{NESM1,NESM2}.
Finite parameter $w^2$, which breaks conformal symmetry, precluding
a conformal transformation connecting the two metrics of the theory,
results from finite source density for neutral
gauge boson field $Z_\mu$, due to time-dependent Ricci scalar $R$ and
finite Higgs scalar field $\phi_0^2$ \cite{NESM2}. Bare field $\Phi_0$
is dressed by the resulting finite but extremely weak $Z_\mu$ field.
This suggests a similar dynamical model for parameter $\lambda$
\cite{NESM4}.  Well-determined values of $R$, $w^2$ \cite{NESM1}
and of $\phi_0^2$ imply negative $\lambda$, incompatible with a massive
Higgs particle.  Observed Hubble expansion for redshifts $z<1$ is
consistent with this inference \cite{NESM2}.

\section{Gravitational field equations}
\par Gravitational field equations are determined by metric
functional derivative
$X^{\mu\nu}= \frac{1}{\sqrt{-g}}\frac{\delta I}{\delta g_{\mu\nu}}$,
where $g$ is the determinant of $g_{\mu\nu}$, and action integral
$I=\int d^4x \sqrt{-g}\sum_a{\cal L}_a$ \cite{WEI72,NES03,MAN06}.
Given $\delta{\cal L}=x^{\mu\nu}\delta g_{\mu\nu}$,
$X^{\mu\nu}=x^{\mu\nu}+\half{\cal L}g^{\mu\nu}$.
Scalar Lagrangian density ${\cal L}_a$ determines variational
energy-momentum tensor $\Theta_a^{\mu\nu}=-2X_a^{\mu\nu}$, evaluated
for a solution of the coupled field equations.  For a bare conformal
field, trace $g_{\mu\nu}X_a^{\mu\nu}=0$ \cite{MAN06}.  
Generalized Einstein equation $\sum_aX_a^{\mu\nu}=0$ is expressed as
$X_g^{\mu\nu}=\half\sum_{a\neq g}\Theta_a^{\mu\nu}$.  Summed trace
$\sum_ag_{\mu\nu}X_a^{\mu\nu}$ vanishes for exact field solutions. 
\par A conformally invariant action integral is defined by Lagrangian
density
${\cal L}_W=-\alpha_g C_\lambda^{\mu\kappa\nu}C^\lambda_{\mu\kappa\nu}$
for Weyl tensor $C_\lambda^{\mu\kappa\nu}$, a traceless projection of
the Riemann tensor \cite{WEY18}.  
After removing a 4-divergence \cite{MAN06},
${\cal L}_g=-2\alpha_g( R^{\mu\nu}R_{\mu\nu}-\thrd R^2 )$, where
$R=g_{\mu\nu}R^{\mu\nu}$.
Conformal
symmetry fixes the relative coefficient of the two quadratic terms. 
For uniform density ${\bar\rho}$ the Weyl tensor vanishes identically, 
so that $X_g^{\mu\nu}\equiv 0$ for a uniform, isotropic cosmos. 
\par Observed excessive galactic rotational velocities have been
studied and parametrized using conformal Weyl Lagrangian density 
${\cal L}_g$ \cite{MAN06,MAN12}.  Hubble expansion has been 
parametrized using conformal Higgs scalar field Lagrangian
density ${\cal L}_\Phi$ \cite{NESM1}.  The generalized Einstein 
equation exactly cancels any vacuum energy density.
\par Metric tensor $g_{\mu\nu}$ is determined by conformal field 
equations derived from ${\cal L}_g+{\cal L}_\Phi$ \cite{NESM3},
driven by energy-momentum tensor $\Theta_m^{\mu\nu}$, where
subscript $m$ refers to conventional matter and radiation.
The gravitational field equation within halo radius $r_H$ is
\begin{eqnarray}
X_g^{\mu\nu}+X_\Phi^{\mu\nu}=\half\Theta_m^{\mu\nu}.
\end{eqnarray}
Defining mean density ${\bar\rho}_G$ and
residual density ${\hat\rho}_G=\rho_G-{\bar\rho}_G$,
and assuming
$\Theta_m^{\mu\nu}(\rho)\simeq
 \Theta_m^{\mu\nu}({\bar\rho})+\Theta_m^{\mu\nu}({\hat\rho})$,
solutions for $r\leq r_G$ of the two equations
\begin{eqnarray} \label{Twoeqs}
X_g^{\mu\nu}=\half\Theta_m^{\mu\nu}({\hat\rho}_G),
X_\Phi^{\mu\nu}=\half\Theta_m^{\mu\nu}({\bar\rho}_G)
\end{eqnarray}
decouple and imply a solution of the full equation.
\par The $X_\Phi$ equation with source density ${\bar\rho}$ has an
exact solution for uniform, isotropic geometry \cite{NESM1,NES13}
in the Robertson-Walker (FLRW) metric, 
\begin{eqnarray} \label{RWmet}
ds^2_{RW}=-dt^2+a^2(t)(\frac{dr^2}{1-kr^2}+r^2d\omega^2),
\end{eqnarray}
where $d\omega^2=d\theta^2+\sin^2\theta d\phi^2$.
This removes any mean density source from the $X_g$ equation, leaving
only the residual density, which integrates to zero over a closed
volume.  There is no residual vacuum energy.
\par Solution of the $X_g$ equation is discussed below for
a static spherical galactic model in Schwarzschild metric 
\begin{eqnarray} \label{ESmet}
ds^2_{ES}=-B(r)dt^2+\frac{dr^2}{B(r)}+r^2d\omega^2.
\end{eqnarray}
Broken conformal symmetry requires a composite hybrid metric such as
\begin{eqnarray} \label{xmet}
ds^2=-B(r)dt^2+a^2(t)(\frac{dr^2}{B(r)}+r^2d\omega^2). 
\end{eqnarray}
Solutions in the two distinct primitive metrics can be made 
consistent by fitting parameters to boundary conditions and setting 
cosmic curvature constant $k=0$, justified by currently observed data. 
 
\section{Conformal gravity}
\par The unique Lagrangian density ${\cal L}_g$ of conformal gravity
theory, constructed from the conformal Weyl 
tensor \cite{WEY18,MAK89,MAN06}, determines Schwarzschild 
gravitational potential
\begin{eqnarray} 
B(r)=-2\beta/r+\alpha+\gamma r-\kappa r^2, 
\end{eqnarray}
valid outside a spherically symmetric
mass/energy source density \cite{MAK89,MAN91}.
Anomalous rotation velocities for 138 galaxies are fitted using
only four universal parameters $\beta^*,\gamma^*,\gamma_0,\kappa$
\cite{MAN06,MAN12,MAO11,OAM15} such that
$\beta=N^*\beta^*=GM/c^2,\gamma=\gamma_0+N^*\gamma^*$.
$N^*$ is galactic baryonic mass $M$ in solar mass units.
Inferred parameter values \cite{MAN06,MAO12}, 
\begin{eqnarray}
\beta^*=1.475\times 10^3 m,
\gamma_0=3.06\times 10^{-28}/m, \nonumber\\
\gamma^*=5.42\times 10^{-39}/m,
\kappa  =9.54\times 10^{-50}/m^2,
\end{eqnarray}
fit conformal gravity to galactic rotation velocities.
\par The static exterior Schwarzschild (ES) metric is defined by
$ds^2_{ES}=-B(r)dt^2+\frac{dr^2}{B(r)}+r^2d\omega^2$ \cite{MAK89},
where $c=\hbar=1$ and $d\omega^2=d\theta^2+\sin^2\theta d\phi^2$.
Given mass/energy source density $f(r)$ enclosed within $\R$,
the field equation in the ES metric is \cite{MAK89,MAN91}
\begin{eqnarray} \label{Beq}
 \dar^4(rB(r))=rf(r), 
\end{eqnarray}
for $f(r)\sim(\Theta_0^0-\Theta_r^r)_m$ determined by source 
energy-momentum tensor $\Theta^{\mu\nu}_m$\cite{MAN06}.

\par
For constants related by $\alpha^2=1-6\beta\gamma$ \cite{MAK91},
\begin{eqnarray}
 y_0(r)=rB(r)=-2\beta+\alpha r+\gamma r^2-\kappa r^3 ,
\end{eqnarray}
is a solution of the tensorial field equation for
source-free $r\geq\R$ \cite{MAK89,MAN91}. 
Derivative functions $y_i(r) =\partial^i_r(rB(r)),0\leq i\leq3$
satisfy differential equations 
\begin{eqnarray}
 \dar y_i=y_{i+1},0\leq i\leq2, \nonumber\\
 \dar y_3=rf(r) .
\end{eqnarray}
The general solution, for
independent constants $c_i=y_i(0)$, determines coefficients
$\beta,\alpha,\gamma,\kappa$ such that at endpoint $\R$
\begin{eqnarray}
 y_0(\R)=-2\beta+\alpha \R+\gamma \R^2-\kappa \R^3,\nonumber\\
 y_1(\R)=\alpha+2\gamma \R-3\kappa \R^2,\nonumber\\
 y_2(\R)=2\gamma-6\kappa \R,\nonumber\\
 y_3(\R)=-6\kappa .
\end{eqnarray}
Gravitational potential $B(r)$ is required to be
differentiable and free of singularities. 
$c_0=0$ prevents a singularity at the origin.  Values 
of $c_1,c_2,c_3$ can be chosen to match outer boundary conditions
$\alpha=1,\gamma=0,\kappa=0$ at $r=\R$. 
This determines parameter $\beta$.
Specific values of $\gamma$ and $\kappa$, consistent with Hubble 
expansion and the observed galactic dark halo \cite{NESM1,NESM3}, 
can be fitted by adjusting
$c_1,c_2,c_3$, subject to $c_0=0,\alpha^2=1-6\beta\gamma$.
\par The proposed particular solution, given $\gamma,\kappa$, is
\begin{eqnarray} \label{rBeq2} 
 rB(r)=y_0(r)=-\sxth\int_0^rq^4fdq+\alpha r-\half r\int_r^\R q^3fdq
\nonumber\\
 +\gamma r^2+\half r^2\int_r^\R q^2fdq
 -\kappa r^3-\sxth r^3\int_r^\R qfdq .
\end{eqnarray}
Integrated parameters $c_i=y_i(0)$ are
$c_1=\alpha-\half \int_0^\R q^3fdq$,
$c_2=2\gamma+\int_0^\R q^2fdq$,
$c_3=-6\kappa-\int_0^\R qfdq$, and at $r=\R$,
$2\beta=\sxth\int_0^\R q^4fdq$.
Term $\gamma r^2+\half r^2\int_r^\R q^2fdq$ in this solution
differs from prior reference\cite{MAK89}.  Here $\gamma$ is a free
parameter that determines generally nonzero $c_2$.
\par Exact solution of the $X_\Phi$ equation with mean source density
${\bar\rho}$ forces the $X_g$ source density to be residual
${\hat\rho}$ or the implied ${\hat f}$.  Because $q^2{\hat f}$
integrates to zero, parameter $\gamma$ cannot depend on 
residual density ${\hat f}$ for any enclosed
spherical source, whether nucleon, star, or galaxy. 
Hence $\gamma$ can only be determined by the $X_\Phi$ equation.
Because $q^2{\hat f}$ integrates to zero,
integrand $q^4{\hat f}$ for Schwarzschild parameter
$\beta$ may involve significant cancellation of positive and
negative terms, contributing to the small magnitude of the
classical gravitational constant.
\par This is compatible with $\gamma$
determined by the conformal Higgs model \cite{NESM1,NES13,NESM3}.
$\kappa$ can be chosen to satisfy boundary condition 
$v^2(r_H)/r_H=\half B'(r_H)c^2=0$ for orbital velocity $v$
at halo radius $r_H$ \cite{MAO11,NESM3}. 

\section{The conformal Higgs model}
\par Assumed nonzero constants $w^2$ and $\lambda$ in Eq.(\ref{DeltaL}) 
are not determined by standard theory\cite{HIG64,CAG98}.
Concurrent solution of the scalar and metric tensor field equations
determines nonzero $\Phi^\dag\Phi=\phi_0^2$ (the Higgs mechanism,
electroweak symmetry-breaking) replacing Einstein
tensor $G^{\mu\nu}$ in the gravitational equation by traceless 
conformal tensor $R^{\mu\nu}-\frth g^{\mu\nu}R$\cite{NESM1,NESM2}. 
\par $\Delta{\cal L}_\Phi$ implies a modified Einstein equation.
A universal nonzero scalar field implies the Higgs mechanism. 
The resulting modified Einstein equation\cite{NESM1,NES13} is
\begin{eqnarray} \label{MEeq}
R^{\mu\nu}-\frth g^{\mu\nu}R+g^{\mu\nu}{\bar\Lambda}=
{\bar\tau}\Theta_m^{\mu\nu}({\bar\rho}).
\end{eqnarray} 
Coefficient ${\bar\tau}$ and cosmological constant ${\bar\Lambda}$
are determined by Higgs model parameters in ${\cal L}_\Phi$.
Because ${\bar\tau}$ depends on the currently unknown prefactor of
${\cal L}_\Phi$, it must be inferred from empirical data.
$\Theta_m^{\mu\nu}$ is the nongravitational energy-momentum tensor.
${\cal L}_g$, constructed from the Weyl 
tensor \cite{WEY18}, vanishes identically in uniform, isotropic 
geometry \cite{MAN06}.  Hubble expansion, described in this
geometry, must be determined by Eq.(\ref{MEeq}).
\par In uniform, isotropic geometry with uniform mass/energy density
${\bar\rho}$, Eq.(\ref{MEeq}) implies a modified Friedmann
equation \cite{NESM1,NES13} for cosmic scale factor $a(t)$,
with $a(t_0)=1$ at present time $t_0$:
\begin{eqnarray} \label{MFeq}
 \frac{{\dot a}^2}{a^2}+\frac{k}{a^2}-\frac{\ddot a}{a}
    =\frac{2}{3}({\bar\Lambda}+{\bar\tau}c^2{\bar\rho}(t)).
\end{eqnarray}
\par Evaluated at time $t$, Ricci scalar $R(t)=
 \frac{{\ddot a}}{a}+\frac{{\dot a}^2}{a^2}+\frac{k}{{\dot a}^2}$.
Neglecting cosmic curvature $k$ in accord with observed data,
conformal sum rule $\Omega_\Lambda(t)+\Omega_m(t)+\Omega_q(t)=1$
follows if conformal Eq.(\ref{MFeq}) is divided by 
${\dot a}^2/a^2$, defining dimensionless Friedmann weights
$\Omega_\Lambda=\frac{2}{3}\frac{{\bar\Lambda}a^2}{{\dot a}^2}$,
$\Omega_m=\frac{2}{3}\frac{{\bar\tau}c^2{\bar\rho}a^2}{{\dot a}^2}$,
and acceleration weight $\Omega_q=\frac{{\ddot a}a}{{\dot a}^2}$  
\cite{NESM1,NESM2}.
Matter and radiation are combined in $\Omega_m$ here while 
$\Omega_\Lambda$ refers to dark energy.  
Hubble function $H(t)=\frac{{\dot a}}{a}(t)\sim [T^{-1}]=h(t)H_0$  
for Hubble constant $H_0=H(t_0)$. Setting $a(t_0)=1$,
$h(t_0)=\frac{{\dot a}}{a}(t_0)=1$ in Hubble units of time $1/H_0$,
length $c/H_0$ and acceleration $H_0^2c/H_0=cH_0$.
\par  For $k=0$ the standard Friedmann equation implies sum rule 
$\Omega_\Lambda(t)+\Omega_m(t)=1$. $\Omega_m=1-\Omega_\Lambda$
requires mass density far greater than observed baryonic mass. 
This has been considered to be a strong argument for dark matter.
With $k=0$ and omitting $\Omega_m$ completely, conformal
sum rule $\Omega_\Lambda(t)+\Omega_q(t)=1$ fits observed data for 
redshifts $z\leq 1$ (7.33Gyr) \cite{NESM1,NESM2}.
This eliminates any need for dark matter to explain Hubble expansion. 
\begin{table}[h]
\caption{Scaled luminosity distance fit to Hubble data} \label{TabA1}
\begin{tabular}{lcccc}
 &                 &          &Theory     &Observed    \\
z& $\Omega_\Lambda$&$\Omega_q$&$H_0d_L/c$Eq.(\ref{Chi_z})
 &$H_0d_L/c$\cite{MAN03}\\ \hline
0.0& 0.732& 0.268& 0.0000& 0.0000\\
0.2& 0.578& 0.422& 0.2254& 0.2265\\
0.4& 0.490& 0.510& 0.5013& 0.5039\\
0.6& 0.434& 0.566& 0.8267& 0.8297\\
0.8& 0.393& 0.607& 1.2003& 1.2026\\
1.0& 0.363& 0.637& 1.6209& 1.6216
\end{tabular}
\end{table}
Luminosity distance $d_L(z)=(1+z)\chi(z)$, for $\Omega_k=0$, is shown
in Table(\ref{TabA1}) for $\alpha=\Omega_\Lambda(t_0)=0.732$,
where\cite{NESM2}
\begin{eqnarray} \label{Chi_z}
\chi(z)=\int_{t(z)}^{t_0}\frac{dt}{a(t)}=
 \int_z^0 dz(1+z)\frac{dt}{dz}=
\nonumber\\
 \int_0^z\frac{dz}{\sqrt{2\alpha\ln(1+z)+1}}.
\end{eqnarray}
Observed redshifts have been fitted to an analytic 
function\cite{MAN03} with
statistical accuracy comparable to the best standard $\Lambda$CDM fit,
with $\Omega_m=0$. Table(\ref{TabA1}) compares CHM $d_L(z)$ to
this Mannheim function.  

\section{Parameters $\gamma$ and $\kappa$}
\par Conformal Friedmann Eq.(\ref{MFeq})\cite{NESM1,NESM2} determines 
cosmic acceleration weight $\Omega_q$, centrifugal back to the 
earliest time\cite{NESM1}. With both weight parameters $\Omega_k$ 
and $\Omega_m$ set to zero, Eq.(\ref{MFeq}) fits scaled 
Hubble function $h(t)=H(t)/H_0$ for redshifts $z\leq 1$ 
as accurately as standard LCDM, with only one free constant.   
This determines Friedmann weights, at present time $t_0$,
$\Omega_\Lambda=0.732, \Omega_q=0.268$ \cite{NESM1}.
Hubble constant $H(t_0)=H_0=2.197\times 10^{-18}/s$ \cite{PLC15}
is independent of these data.
\par Because fixed $H_0$ implies uniform $\Omega_\Lambda$,
the dimensionless sum rule\cite{NESM1} with $\Omega_k=0$ determines
$\Omega_q(\rho_m)=1-\Omega_\Lambda-\Omega_m$ in the cosmic background,  
and $\Omega_q(0)=1-\Omega_\Lambda$ in the depleted halo\cite{NESM3}.   
Observed nonclassical gravitational acceleration $\half\gamma c^2$
in the halo is proportional to 
$\Delta\Omega_q=\Omega_q(0)-\Omega_q(\rho_m)=
\Omega_m(\rho_m)$\cite{NESM3}, where, given $\rho_m$ and $H_0$, 
$\Omega_m(\rho_m)=
\frac{2}{3}\frac{{\bar\tau}c^2\rho_m}{H^2_0}$\cite{NESM1}. 
Thus the depleted halo model determines $\gamma$ from
uniform universal cosmic baryonic mass density $\rho_m/c^2$,
which includes free radiation energy density here.
\par Converted from Hubble units, this implies centripetal
acceleration $\half\gamma c^2=-cH_0\Omega_m(\rho_m)$\cite{NESM3}.
Positive $\rho_m$ implies $\Omega_m<0$ 
because coefficient ${\bar\tau}<0$ \cite{MAN06,NESM1}. Hence 
$\Delta\Omega_q=\Omega_m<0$ is consistent with nonclassical
centripetal acceleration $\half\gamma c^2$, confirmed by inward
deflection of photon geodesics observed in gravitational
lensing \cite{NESM3}.
\par  This logic is equivalent to requiring continuous radial
acceleration across halo boundary $r_H$:
\begin{eqnarray}
\half\gamma c^2-cH_0\Omega_q(0)=-cH_0\Omega_q(\rho_m).
\end{eqnarray}
Signs here follow from the definition of $\Omega_q$ as centrifugal
acceleration weight.
\par The Milky Way is treated here as a typical model 
galaxy\cite{MCG08,OAM15}. Specific model parameters are galactic mass
$M=N^* M_\odot=1.207\times 10^{41}kg$, for
$N^*=6.07\times 10^{10}, M_\odot=1.989\times 10^{30}kg$.
To avoid assuming specific mass dependence, only results dependent
on total $\gamma=\gamma_0+N^*\gamma^*=6.35\times 10^{-28}/m$, are
considered here.  The depleted halo model \cite{NESM3} identifies 
$\kappa$ as a cutoff parameter at halo radius $r_H=\half\gamma/\kappa$. 
This gives specific empirical values 
$r_H=\half\gamma/\kappa=33.28\times 10^{20}m$,
and cosmic mass density $\rho_m/c^2=7.817\times 10^{-25} kg/m^3$,
using $M=1.207\times 10^{41}kg=\frac{4\pi}{3}r^3_H\rho_m/c^2$.
\par For empirical $\gamma=6.35\times 10^{-28}/m$,
given $\rho_m$ and $H_0$, dimensionless 
$-\half\gamma\frac{c}{H_0}=\Omega_m(\rho_m)$ implies
$\Omega_m= -0.0433$, consistent with observed Hubble 
expansion \cite{NESM1}.
$\Omega_m(\rho_m)=
\frac{2}{3}\frac{{\bar\tau}c^2\rho_m}{H^2_0}=-0.0433$ 
implies nonclassical gravitational constant
\begin{eqnarray}
{\bar\tau}=-4.96\times 10^{-47}s^2/kg/m,
\end{eqnarray}
for the conformal Higgs model \cite{NESM1,NES13}. 
The standard Newton/Einstein gravitational constant is
\begin{eqnarray}
\tau=\frac{8\pi G}{c^4}=2.08\times 10^{-43}s^2/kg/m.
\end{eqnarray}
\par For a single spherical solar mass isolated in a galactic halo,
mean internal mass density ${\bar\rho}_\odot$ within $r_\odot$
determines an exact solution of the conformal Higgs gravitational
equation, giving internal acceleration 
$\Omega_q({\bar\rho}_\odot)$.
\par Given $\gamma$ outside $r_\odot$,
continuous acceleration across boundary $r_\odot$,
\begin{eqnarray}  
\half\gamma_{\odot,in}c^2-cH_0\Omega_q({\bar\rho}_\odot)=
 \half\gamma c^2-cH_0\Omega_q(0),
\end{eqnarray}
determines constant $\gamma_{\odot,in}$ valid inside $r_\odot$.
\par $\gamma_{\odot,in}$ is determined by local mean source density
${\bar\rho}_\odot$. $\gamma$ in the halo is not changed.
Its value is a constant of integration that cannot vary
in the source-free halo.  Hence there is no way to determine
a mass-dependent increment to $\gamma$.  This replaces the usually 
assumed $\gamma=\gamma_0+N^*\gamma^*$ by $\gamma=\gamma_H$, 
determined at halo boundary $r_H$.

\section{Conformal scalar field: implied $\lambda$}
\par The conformal scalar field equation
including parametrized $\Delta{\cal L}_\Phi$ is \cite{MAN06,NESM1}
\begin{eqnarray}\label{Phieq}
\frac{1}{\sqrt{-g}}\partial_\mu(\sqrt{-g}\partial^\mu\Phi)=
 -(\sxth R-w^2+2\lambda\Phi^\dag\Phi)\Phi.
\end{eqnarray}
For $k=0$, $\sxth R(t)=
 h^2(t)(1+\Omega_q(t))>h^2(t)\Omega_\Lambda=w^2$,
in Hubble units\cite{NESM1} .
\par Ricci scalar $R$ introduces gravitational effects.
Time-dependent $R=6(\xi_0(t)+\xi_1(t))$, where
$\xi_0(t)=\frac{\ddot a}{a}$ and
$\xi_1(t)=\frac{{\dot a}^2}{a^2}+\frac{k}{a^2}$\cite{NESM1}. 
For $k=0$, $\sxth R(t)=
 h^2(t)(2-\Omega_\Lambda(t)-\Omega_m(t))=
 h^2(t)(1+\Omega_q(t))>h^2(t)\Omega_\Lambda(t)))=w^2$.
Hence $\zeta(t)=\sxth R(t)-w^2>0$.
\par Only real-valued solution $\phi(t)$ is relevant in uniform,
isotropic geometry. The field equation is
\begin{eqnarray} \label{phieq}
\frac{{\ddot\phi}}{\phi}+3\frac{{\dot a}}{a}\frac{{\dot\phi}}{\phi}=
 -(\sxth R(t)-w^2+2\lambda\phi^2).
\end{eqnarray}
\par Omitting $R$ and assuming constant $w^2$ and $\lambda>0$, Higgs
solution $\phi_0^2=w^2/2\lambda$ \cite{HIG64} is exact.
All time derivatives drop out.  In the conformal scalar field equation,
cosmological time dependence of Ricci scalar $R(t)$,
determined by the CHM Friedmann cosmic evolution equation,
introduces nonvanishing time derivatives and implies $\lambda<0$.
Time-dependent terms in the scalar field equation can be included in
$w^2(t)=\frac{{\dot\phi}^2}{\phi^2}-\frac{{\ddot\phi}}{\phi}
-3h(t)\frac{{\dot\phi}}{\phi}$.
For $\zeta(t)=\sxth R(t)-w^2(t)>0$, $\phi^2(t)=-\zeta(t)/2\lambda$
is an exact solution of Eq.(\ref{phieq}), but requires $\lambda<0$. 
\par $\zeta>0$ for computed $R(t)$\cite{NESM1} implies $\lambda<0$.
$\hbar\phi(t_0)=174GeV$\cite{AMS08}$=1.203\times 10^{44}\hbar H_0$. 
In Hubble units, for $\Omega_m=0$, $\zeta(t_0)=2\Omega_q(t_0)=0.536$.
For empirical $\phi(t)$, $\lambda(t)=\zeta/(-2\phi^2)$. Given 
$\zeta(t_0)$ and $\phi(t_0)$, 
dimensionless $\lambda(t_0)=-0.185\times 10^{-88}$. 

\section{Radial acceleration and baryonic Tully-Fisher relations}
\par Conformal gravity, the CHM, and the depleted halo model are
consistent with empirical relations inferred from observed galactic 
orbital velocities, dark halos, and accelerating Hubble expansion.
\par Static spherical geometry defines Schwarzschild
potential $B(r)$.  For a test particle in a stable exterior
circular orbit with velocity $v$ the centripetal acceleration is
$a=v^2(r)/r=\half B'(r)c^2$.
Newtonian $B(r)=1-2\beta/r$, where $\beta=GM/c^2$,
so that $a_N=\beta c^2/r^2=GM/r^2$.
\par CG adds nonclassical $\Delta a$ to $a_N$, so that orbital velocity
squared is the sum of $v^2(a_N;r)$ and $v^2(\Delta a;r)$, which
cross with equal and opposite slope at some $r=r_{TF}$.  This defines
a flat range of $v(r)$ centered at stationary point $r_{TF}$, without
constraining behavior at large $r$.
\par MOND\cite{MIL83,SAN10,FAM12} modifies the Newtonian force
law for acceleration below an empirical scale $a_0$.
Using $y=a_N/a_0$ as independent variable,
for assumed universal constant $a_0$,
MOND postulates an interpolation function $\nu(y)$ such that observed
radial acceleration $a=f(a_N)=a_N\nu(y)$.
A flat velocity range approached
asymptotically requires $a^2\to a_Na_0$ as $a_N\to0$.
For $a_N\gg a_0$, $\nu\to1$ and for $a_N\ll a_0$,
$\nu^2\to 1/y$.  This implies asymptotic limit $a^2\to a_0a_N$
for small $a_N$, which translates into an
asymptotically flat galactic velocity function $v(r)$ for
large orbital radius $r$\cite{MIL83}.
For $a_N\ll a_0$, MOND $v^4=a^2r^2\to GMa_0$,
the empirical baryonic Tully-Fisher relation\cite{TAF77,MCG05,MCG11}.
\par In conformal gravity (CG), centripetal acceleration
$a=v^2/r$ determines exterior orbital velocity
$v^2/c^2=ra/c^2=\beta/r+\half\gamma r-\kappa r^2$,
compared with asymptotic $ra_N/c^2=\beta/r$.
If the asymptotic Newtonian function is valid at $r$ and
$2\kappa r/\gamma$ can be neglected, the slope of $v^2(r)$ vanishes at
$r^2_{TF}=2\beta/\gamma$.  This implies that
$v^4(r_{TF})/c^4=(\beta/r_{TF}+\half\gamma r_{TF})^2=
  2\beta\gamma$\cite{MAN97}.
\par This is the Tully-Fisher relation, exact at stationary point
$r_{TF}$ of the $v(r)$ function.  Given $\beta=GM/c^2$,
$v^4=2GM\gamma c^2$, for relatively constant $v(r)$ centered
at $r_{TF}$.  For comparison, MOND $v^4=GMa_0$ would be identical
if $a_0=2\gamma c^2$\cite{MAN97}, for mass-independent $\gamma$.
\par McGaugh et al\cite{MLS16} have recently shown for 153 disk
galaxies that observed radial acceleration $a$ is effectively a 
universal function of the expected classical Newtonian acceleration 
$a_N$, computed for the observed baryonic distribution. Galactic mass 
is determined directly by observation, removing uncertainty due to 
adjustment of mass-to-light ratios for individual 
galaxies in earlier studies.
\par The existence of such a universal correlation function,
$a(a_N)=a_N\nu(a_N/a_0)$ is a basic postulate of
MOND\cite{MIL83,FAM12}. CG implies a similar correlation function if 
nonclassical parameter $\gamma$ is mass-independent\cite{NESM7}.
\par Outside an assumed spherical source mass, conformal Schwarzschild
potential $B(r)$ determines circular geodesics such that
$v^2/c^2=ra/c^2=\half rB^\prime(r)=\beta/r+\half\gamma r-\kappa r^2$.
The Kepler formula is $ra_N/c^2=\beta/r$.
Well inside a galactic halo boundary, $2\kappa r/\gamma$ can be
neglected.  This defines correlation function $f(a_N)=a_N+\Delta a$,
if $\Delta a=\half\gamma c^2$ is a universal constant \cite{NESM7}, 
which would be true if Mannheim acceleration parameter $\gamma$ were 
independent of galactic mass.  This contradicts the definition
$\gamma=\gamma_0+N^*\gamma^*$ used in fitting rotation data for 138
galaxies to conformal gravity \cite{MAN06,MAO12}. 
$\gamma\simeq\gamma_H$, a universal constant, as shown here,
could be tested by fitting data for known galactic mass \cite{MLS16}
to conformal gravity, to determine empirical $\gamma(M)$.
\par CG implies mass-dependent effects at large radii, not described
by MOND.  CG orbital velocity drops to zero at an outer boundary
(dark halo radius)\cite{MAO11,NESM3}).
An empirical distinction between MOND and CG requires accurate
rotational data at large galactic radii\cite{NESM7}.  
CG parameter $\kappa$, inferred from such data\cite{MAO11}, and 
consistent with the boundary radius inherent in the depleted halo 
model\cite{NESM3}, does not have a counterpart in MOND.

\section{Qualitative results relevant to physical data}
\par For nonclassical acceleration parameter $\gamma$, Ricci 
scalar $R$ is in general singular as $r\to0$, contrary to a constraint 
imposed implicitly by previous analyses \cite{MAK94,FLA06,BAV09,YOO13}.
Removing this constraint allows
$\gamma$ to be determined independently by the conformal
Higgs model of Hubble expansion, as shown here.  Parameter $\kappa$ 
determines a cutoff at the galactic halo boundary, consistent with
observed rotational velocities and galactic lensing.
Singular $R$ as $r\to0$ is consistent with occurrence of supermassive 
black holes at galactic centers.
\par The classical Schwarzschild potential equation
has two independent parameters.  One is fixed by requiring regularity
at the origin, while the other determines the Newtonian radial
potential, which extends to infinity.  The two additional parameters 
of the fourth-order conformal potential allow consistency with  
the conformal Higgs field equation while 
terminating radial acceleration at a halo radius.  
This implies physically that a static galactic gravitational field does 
not extend beyond its halo.  The classical second-order 
equation has no such cutoff.
The hybrid metric considered here specifically sets 
cosmic curvature $k=0$, consistent with the current empirical estimate, 
removing the longstanding flatness paradox. 
\par The exterior Schwarzschild solution for conformal gravity is valid
on a subatomic scale.  Mass/energy densities $\rho_n$ and $\rho_p$
within a neutron and proton, respectively, must certainly differ.
Assuming spherical symmetry, $r^4$-weighted integrals
$\beta_n,\beta_p$ and $r^2$-weighted mass integrals $m_n,m_p$
cannot be expected to be proportional with the same constant.
Two constants $G_n/c^2=\beta_n/m_n, G_p/c^2=\beta_p/m_p$ are defined.
\par Schwarzschild parameters
$\beta_n$ and $\beta_p$, for neutron and proton, respectively,
are not necessarily proportional to the particle mass.
For a macroscopic gravitating object, the effective gravitational
constant is $G_{eff}/c^2=(N_n\beta_n+N_p\beta_p)/(N_n m_n+N_p m_p)$,
which depends on the ratio $N_n/N_p$, plus a small effect of electronic
mass and net chemical energy. This may explain the persistent problem
of defining an accurate value of $G$\cite{SCH14,RAL14}.
\par Mannheim parameter $\kappa$\cite{MAN06,MAO11,OAM15} is shown here
to be consistent with the outer halo boundary radius postulated in the
depleted halo model \cite{NESM3}. Vanishing centripetal acceleration 
and consequent elimination of stable galactic orbits beyond a halo 
radius is a unique feature of conformal theory. It cannot be 
understood in terms of dark matter or of locally modified Newtonian 
dynamics (MOND) \cite{MIL83,SAN10,FAM12}.  If a static gravitational 
potential cannot extend beyond this boundary, galactic interactions 
require colliding halos. Unbounded field energy due to a source of 
infinite extent is ruled out.
\par Well-defined values of relevant parameters here conflict with
determination of a massive Higgs boson by the conformal Higgs model,
although the Higgs mechanism, requiring a self-generated finite
$\phi_0$ field amplitude, is confirmed.  This apparent conflict with
interpretation of the recently observed LHC 125GeV resonance
\cite{ATL12,CMS12} may be resolved if the observed object differs from 
the anticipated Higgs boson.  A candidate to be considered can be shown 
to imply the correct empirical value of Higgs $\lambda$\cite{NESM4}.

\section{Conclusions on formalism}
\par The gravitational field equation deduced from conformal Weyl
theory for defined source density $f(r)$ in the static spherically
symmetric Schwarzschild metric is
$\dar^4(rB(r))=rf(r)$ \cite{MAK89,MAN91,MAN06}.
Exterior Schwarzschild potential function 
$B(r)=-2\beta/r+\alpha+\gamma r-\kappa r^2$
is exact outside any enclosed mass/energy radial source density.
There are four independent parameters, of which only two
have specified values: 
$c_0=[rB(r)]_{r\to0}=0,\alpha^2=1-6\beta\gamma$.
For arbitrary radial source density a particular solution exists
that determines $\beta$ for any given $\gamma$ and $\kappa$.
\par Consistency with conformal theory of observed Hubble expansion
\cite{NES13} and dark galactic halos\cite{NESM3} removes uniform
mean energy-momentum source density from the conformal gravitational
field equation for an extended bounded volume.  This implies that
nonclassical acceleration parameter $\gamma$ is determined by
the purely time-dependent modified Friedmann equation.  
$\gamma$ due to the depleted dark
halo model\cite{NESM3} has the expected magnitude, consistent
with negligible effect at subgalactic distances.  
Except for small effects of chemical composition, this removes any
conflict between classical and conformal gravity for subgalactic
phenomena. At close interstellar distances,
conformal theory predicts no measurable deviation from standard
general relativity.  These conclusions resolve objections raised by
previous authors \cite{FLA06,BAV09,YOO13,KH016}.
\par A recent study of galactic rotation data designed to eliminate 
mass adjustment of individual galaxies\cite{MLS16} indicates that 
empirical galactic $\gamma$ may be mass-independent\cite{NESM7}.
If so, derivations given here are consistent with this study
of rotational velocities and with MOND constant 
$a_0$\cite{MIL83,SAN10,FAM12}.  Empirical $\gamma$ 
should be optimized for the data of Ref\cite{MLS16}.
\par The present analysis, together with earlier studies of Hubble
expansion and dark halos\cite{NESM1,NES13,NESM2,NESM3}, establishes
relationships among parameters limited to observed data for individual
galaxies. The interdependence of CG and CHM shown by the present 
analysis indicates that extension of accurate CHM results beyond 
redshift unity requires simultaneous integration of functions 
$\phi(t)$ and $a(t)$.  Extension to galactic growth, interactions, 
and clusters has not yet been explored.
\par Basic parameters inferred here from empirical galactic rotation
data \cite{MAN97,MAN06,MAO11,MAO12,OAM12,OAM15} include:\\
Cosmic mass density $\rho_m/c^2=7.82\times 10^{-25}kg/m^3$,\\
Friedmann mass weight $\Omega_m=-4.33\times 10^{-2}$,\\
$X_\Phi$ gravitational constant
 ${\bar\tau}=-4.96\times 10^{-47}s^2/kg/m$,\\
MOND $a_0=2\gamma c^2=1.14\times 10^{-10}m/s^2$.\\
Milky Way halo radius $r_H=107.8 kpc$,\\
Milky Way Tully-Fisher radius $r_{TF}=17.2 kpc$,\\
compared with empirical galactic radius $r_G\simeq 15.0 kpc$.
\par The author is indebted to Prof. Keith Horne for helpful
questions and comments.

\end{document}